\documentclass[longauth]{aa} 
\usepackage{natbib}
\usepackage{txfonts}
\usepackage[colorlinks,citecolor=blue]{hyperref}%
\usepackage{graphicx}
\usepackage{txfonts}

\usepackage{changes}

\newcommand{\orcid}[1]{\unskip\protect\href{https://orcid.org/#1}{\protect\includegraphics[width=8pt,clip]{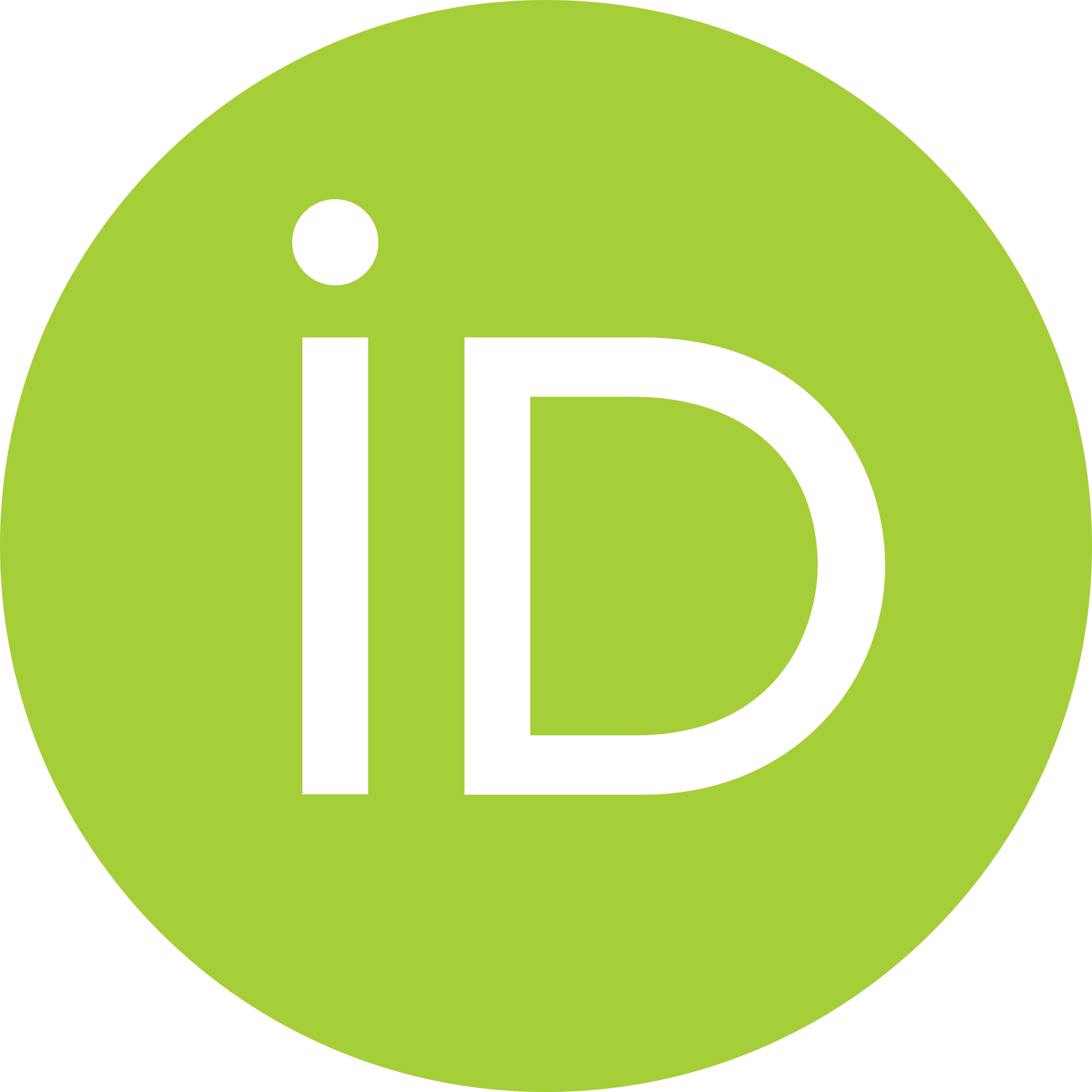}}}

\newcommand{\varcsec}{^{\prime\!\hskip0.6pt\prime}}
\begin{document} 

   \title{Wavefront error of PHI/HRT on Solar Orbiter at various heliocentric distances }

  \author{F.~Kahil\inst{1}\thanks{\hbox{Corresponding author: F.~Kahil.} \hbox{\email{kahil@mps.mpg.de}}}\orcid{0000-0002-4796-9527}
     \and
   A.~Gandorfer\inst{1}\orcid{0000-0002-9972-9840} \and
   J.~Hirzberger\inst{1} \and
   D.~Calchetti\inst{1}\orcid{0000-0003-2755-5295} \and
   J.~Sinjan\inst{1}\orcid{0000-0002-5387-636X} \and   
   G.~Valori\inst{1}\orcid{0000-0001-7809-0067} \and
   S.K.~Solanki\inst{1}\orcid{0000-0002-3418-8449} \and
   M.~Van~Noort\inst{1} \and
   K.~Albert\inst{1}\orcid{0000-0002-3776-9548} \and
   N. Albelo~Jorge\inst{1} \and
   A.~Alvarez-Herrero\inst{2}\orcid{0000-0001-9228-3412}  \and
   T.~Appourchaux\inst{3}\orcid{0000-0002-1790-1951} \and 
   L.R.~Bellot~Rubio\inst{4} \orcid{0000-0001-8669-8857}\and
   J.~Blanco~Rodr\'\i guez\inst{5}\orcid{0000-0002-2055-441X} \and
   A.~Feller\inst{1} \and   
   B.~Fiethe\inst{6}\orcid{0000-0002-7915-6723} \and
   D.~Germerott\inst{1} \and
   L.~Gizon\inst{1,10} \orcid{0000-0001-7696-8665} \and 
   L.~Guerrero\inst{1} \and
   P.~Gutierrez-Marques\inst{1}\orcid{0000-0003-2797-0392} \and
   M.~Kolleck\inst{1} \and
   A.~Korpi-Lagg\inst{1}\orcid{0000-0003-1459-7074} \and
   H.~Michalik\inst{6} \and
   A.~Moreno~Vacas\inst{4}\orcid{0000-0002-7336-0926} \and
   D.~Orozco~Su\' arez\inst{4}\orcid{0000-0001-8829-1938} \and
   I.~P\' erez-Grande\inst{8}\orcid{0000-0002-7145-2835} \and 
   E.~Sanchis Kilders\inst{5}\orcid{0000-0002-4208-3575} \and
   J.~Schou\inst{1}\orcid{0000-0002-2391-6156} \and
   U.~Sch\" uhle\inst{1}\orcid{0000-0001-6060-9078} \and
   J.~Staub\inst{1}\orcid{0000-0001-9358-5834} \and
   H.~Strecker\inst{4}\orcid{0000-0003-1483-4535} \and
   J.C.~del~Toro~Iniesta\inst{4}\orcid{0000-0002-3387-026X}\and
   R.~Volkmer\inst{9}\and
   J.~Woch\inst{1}\orcid{0000-0001-5833-3738}
   }

   \institute{
         Max-Planck-Institut f\"ur Sonnensystemforschung, Justus-von-Liebig-Weg 3,
         37077 G\"ottingen, Germany \\ \email{solanki@mps.mpg.de}
         \and
         Instituto Nacional de T\' ecnica Aeroespacial, Carretera de
         Ajalvir, km 4, E-28850 Torrej\' on de Ardoz, Spain
         \and
         Univ. Paris-Sud, Institut d’Astrophysique Spatiale, UMR 8617,
         CNRS, B\^ atiment 121, 91405 Orsay Cedex, France
         \and
         Instituto de Astrofísica de Andalucía (IAA-CSIC), Apartado de Correos 3004,
         E-18080 Granada, Spain \\ \email{jti@iaa.es}
         \and
         Universitat de Val\`encia, Catedr\'atico Jos\'e Beltr\'an 2, E-46980 Paterna-Valencia, Spain
         \and
         Institut f\"ur Datentechnik und Kommunikationsnetze der TU
         Braunschweig, Hans-Sommer-Str. 66, 38106 Braunschweig,
         Germany
         \and
         Instituto Universitario "Ignacio da Riva", Universidad Polit\'ecnica de Madrid, IDR/UPM, Plaza Cardenal Cisneros 3, E-28040 Madrid, Spain
         \and
         Leibniz-Institut für Sonnenphysik, Sch\" oneckstr. 6, D-79104 Freiburg, Germany
         \and
         Institut f\"ur Astrophysik, Georg-August-Universit\"at G\"ottingen, Friedrich-Hund-Platz 1, 37077 G\"ottingen, Germany}

 \abstract
   {} 
   { We use wavefront sensing to characterise the image quality of the the High Resolution Telescope (HRT) of the Polarimetric and Helioseismic Imager (SO/PHI) data products during the second remote sensing window of the Solar Orbiter's (SO) nominal mission phase. The ultimate aims are to reconstruct the HRT data by deconvolving with the HRT point spread function (PSF) as well as to correct just for the effects of optical aberrations on the data.   }
   {We use a pair of focused-defocused images to compute the wavefront error and derive the PSF of HRT by means of a phase diversity (PD) analysis. }
   {The wavefront error of HRT depends on the orbital distance of SO to the Sun. At distances $>0.5$\,au, the wavefront error is small, and stems dominantly from the inherent optical properties of HRT. At distances $<0.5$\,au the thermo-optical effect of the Heat Rejection Entrance Window (HREW) becomes noticeable. We develop an interpolation scheme for the wavefront error which depends on the thermal variation of the HREW with the distance of SO to the Sun. We also introduce a new level of image reconstruction, termed "aberration correction", which aims to reduce the noise caused by image deconvolution, while removing the aberrations caused by the HREW. }
   {The computed PSF via phase diversity reduces significantly the degradation caused by the HREW in the near-perihelion HRT data. In addition, the aberration correction increases the noise by a factor of only $1.45$ compared to the factor of $3$ increase which results from the usual PD reconstructions. }
   \keywords{Sun: photosphere, magnetic field, continuum -- Techniques: Polarimetry, Imagery, Phase diversity, wavefront error, PSF, MTF, Zernike, noise -- Instrumentation: SO, PHI, HRT}
         
   \maketitle

\section{Introduction}

\begin{figure*}[!h]
\centering
\includegraphics[width=\textwidth]{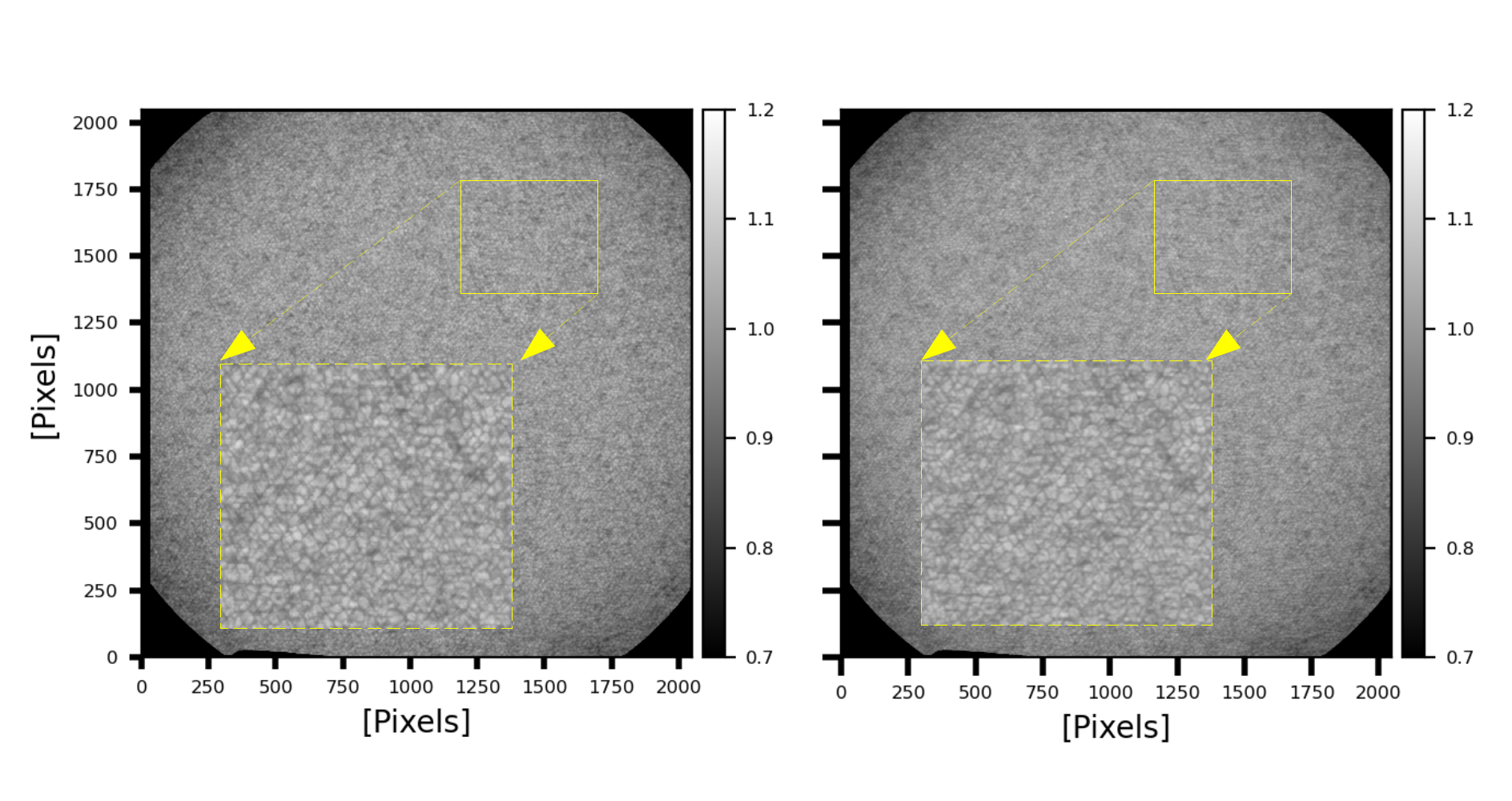}
  \caption{The HRT PD image pair of 22 March 2022 at $0.334$\,au. The focused image is shown to the left, while the defocused image (by half a wavelength with respect to the focused image) is shown to the right. We draw in yellow a blow-up of the same region in both images.  }
  
     \label{pd_pair_march}
\end{figure*}
\begin{figure}[!h]
\centering
\includegraphics[width=\columnwidth]{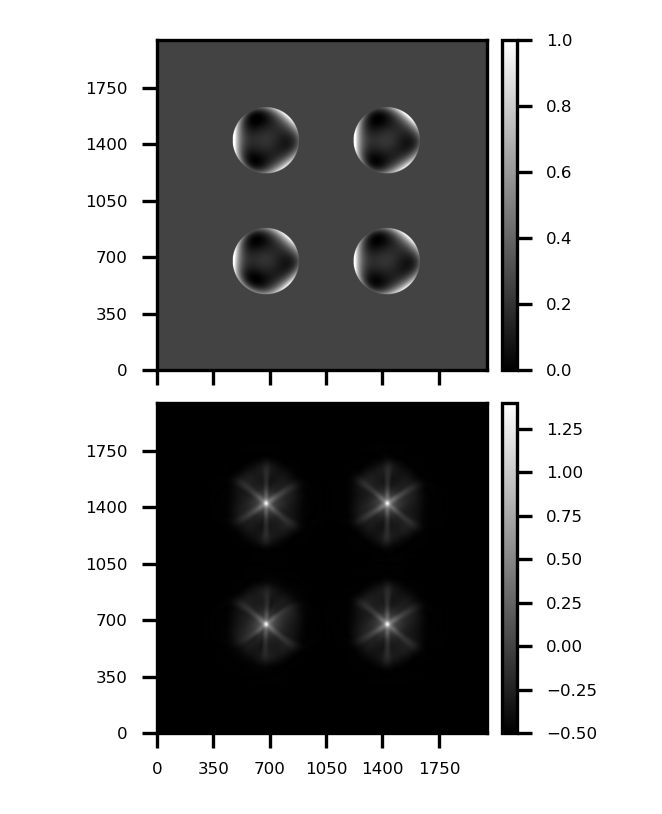}
  \caption{Results of the PD analysis. The wavefront error (upper panel, in units of wavelength) and Modulation Transfer Function (MTF, lower panel) in four sub-regions, of $750\times 750$ pixels each, of the entire FOV of HRT ($2048\times2048$ pixels) .
  }
  
     \label{wfs_mtfs_march}
\end{figure}

\begin{figure*}[!h]
\centering
\includegraphics[width=\textwidth]{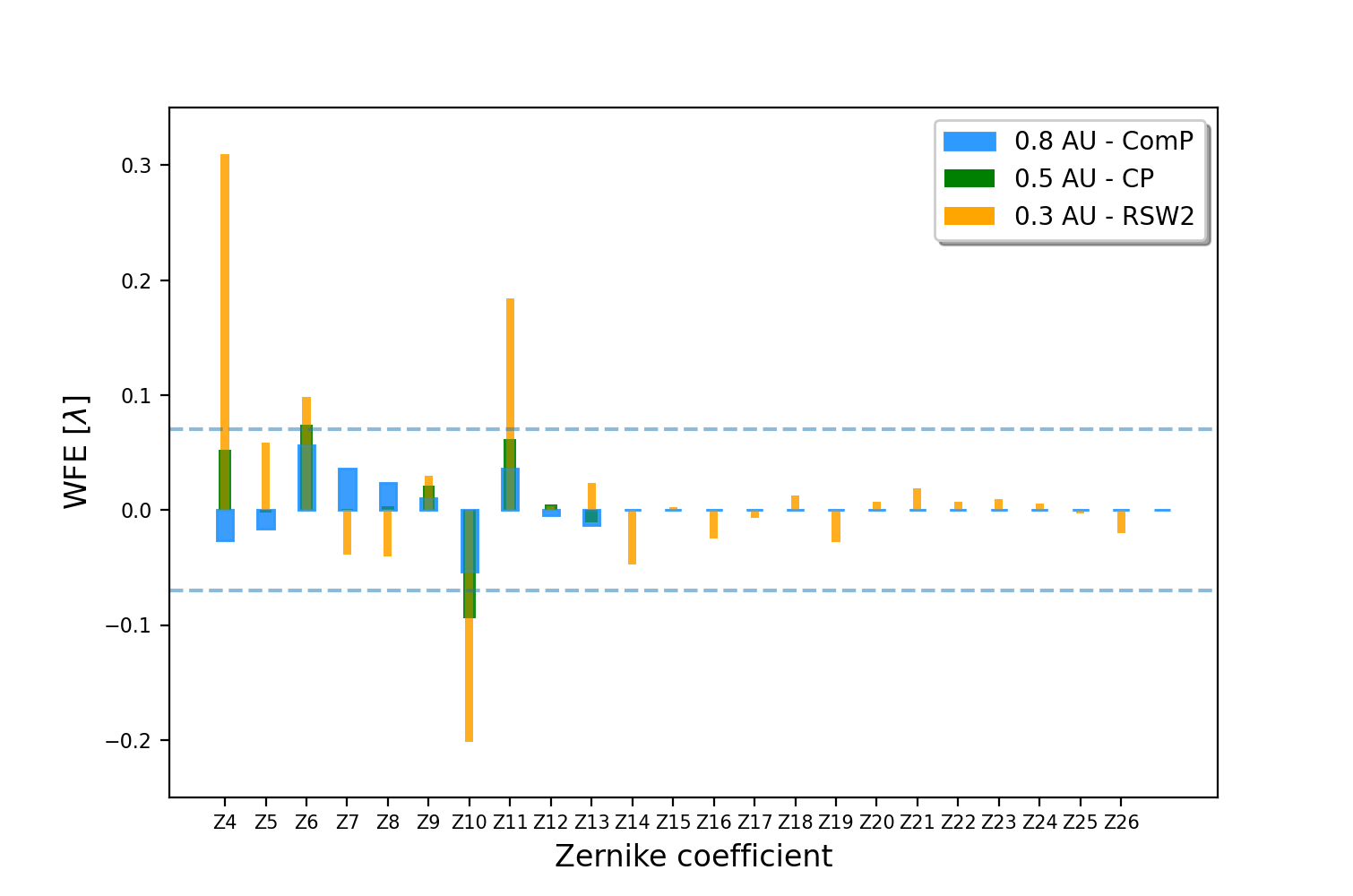}
  \caption{The Zernike coefficients distribution as deduced from the analysis of the PD datasets for each orbit. The dashed blue lines indicate the RMS WFE values ($\pm \lambda/14$) which correspond to a diffraction limited performance.}
  
     \label{all_zernikes}
\end{figure*}

Solar Orbiter \citep[SO,][]{muller_solar_2020} entered its low-orbital nominal mission phase \citep[NMP,][]{zouganelis_solar_2020} in late November 2021. During the first orbit of the NMP most of the observations by the remote sensing instruments were carried out in three remote sensing windows (RSWs) spanning the period of  01 March 2022 until 06 April 2022. The closest approach of SO to the Sun during these RSWs, of $0.32$\,au, was reached on 26 March 2022. Among the remote sensing instruments on-board, here we consider the Polarimetric and Helioseismic Imager \citep[SO/PHI,][]{solanki2020}, which provides measurements of the magnetic field, either of the full solar disc, or at higher resolution of a small portion of the solar surface. 

The latter observations are carried out with the High Resolution Telescope of SO/PHI \citep[HRT, see][]{gandorfer18}, which is a two-mirror system with a decentered Ritchey-Chr\'etien configuration. The entrance aperture of the telescope has a diameter of $140$\,mm. With an effective focal length of $4125$\,mm in the focal plane, the angular sampling corresponding to a working wavelength of $\lambda = 6173$\,\AA{} is $0.5\varcsec$. This angular sampling equals about $100$~km on the solar surface at the closest perihelion of SO at $0.28$\,au. 

Changes in the very high image quality achieved by HRT are driven mainly by the thermal environment. In particular, the Heat Rejection Entrance Window \citep[HREW,][]{solanki2020} acts as a passive thermal element in the heat-shield assembly of the spacecraft, and exhibits a large temperature variation along the highly elliptic orbit of Solar Orbiter. The HREW is designed such that the temperature gradient across the glass plates of the window is radially symmetric. Thus, the produced thermal lensing effect (the dependence of the refractive index of the glass on the temperature) introduces only a defocus term which can be compensated for by the HRT Refocus Mechanism \citep[HRM, see also][]{solanki2020}. The amplitude of this gradient is estimated to produce a defocus up to 4\,$\lambda$ at perihelion (where the glass temperature reaches about 200\,$^\circ$C in the center of the window with a 20\,$^\circ$C radial gradient towards the outer edges).

These conditions, however, are not perfectly fulfilled in flight. Therefore, it is expected that, at close solar proximity, the wavefront error (WFE) is compromised by higher order residual optical aberrations, which the HRM is incapable of removing. Then again, the HRM enables acquiring a pair of focused and defocused images of the solar scene, which could be used, by means of phase diversity analysis \citep[PD,][]{paxman_1990,loefdahl_scharmer_94, paxman_phase-diversity_1994}, to determine the optical degradations of HRT due to deformations of the HREW. 

PD is a powerful technique which can be employed to capture the low-to-medium order telescope aberrations \citep{gonsalves_fundamentals_1983,gonsalves_phase_1985}. These are usually reflected in the total wavefront error at the exit pupil of a telescope, here HRT, or in the point spread function (PSF) in the corresponding image plane \citep{goodman1996introduction}. Internal optical aberrations such as coma or astigmatism are inherent to the HRT, and originate from imperfections in the complex optical system, mainly thermo-elastic despace errors in the two mirror system \citep{wilson_image_1999}. Residual defocus, spherical aberration or trefoil terms are not expected to originate from the SO/PHI optics and are produced by thermal gradients on the HREW.

During the Near-Earth Commissioning Phase (NECP, $0.8$\,au) and in the second remote sensing checkout window of the Cruise Phase (CP, $0.5$\,au) the optical aberrations of HRT were characterised. The results are published in \cite{kahil_2022_PD}. They found a common wavefront error over the
field of view (FOV) of HRT and that the WFE is larger during CP when SO is closer to the Sun.
From March to April 2022, SO reached, for the first time, solar distances below $0.5$\,au.  We aim in this work to evaluate the image quality of HRT data products taken during the first two remote sensing windows of the NMP. In Section~\ref{pd_analysis} we present the PD data and our approach for fitting the WFE. In Sections~\ref{psf_interpolation} and \ref{aberration_correction} we describe the methods we use to reconstruct the near-perihelion data with the available PD measurements of HRT and show our results. These are discussed along with conclusions in Section~\ref{discussion}. 

\begin{table}[h]
\caption{Summary of phase diversity data. \textit{First column:} The acquisition dates of the PD image pair. \textit{Second column:} The Heliocentric distance of SO. \textit{Third column:} The optimal number of Zernike polynomials in the PD fitting routine. \textit{Fourth column:} The total RMS wavefront error. }
\centering
\begin{tabular}{c c c c c c}
\hline
Date & Distance [au]&$Z$  & RMS WFE [$\lambda$]   \\
\hline
20$-$04$-$2020 &  0.82 &10 & 1/10\\
\hline
20$-$02$-$2021 & 0.523  &  10 & 1/7\\
\hline
22$-$03$-$2022 & 0.334 & 23& 1/2.27\\
\hline 

\end{tabular}

\label{tab1}
\end{table}

\begin{figure}[!h]
\centering
\includegraphics[width=\linewidth]{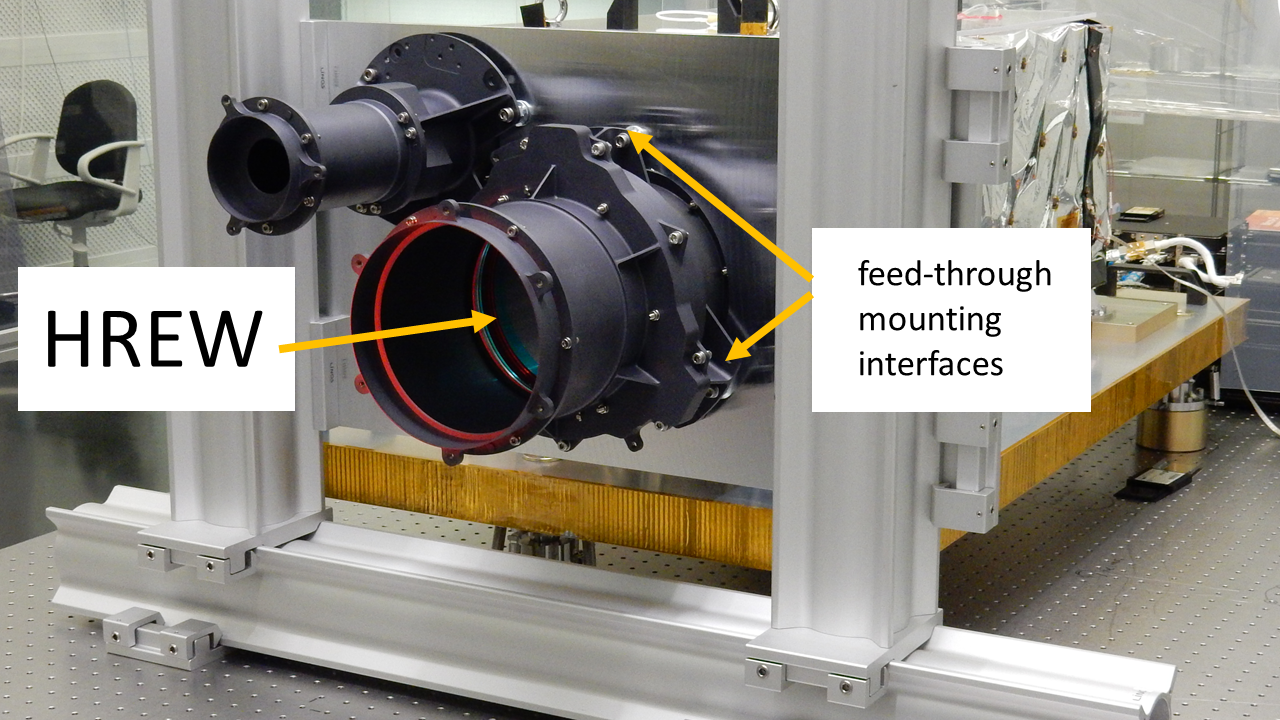}
  \caption{The feed-through containing the HREW during SO/PHI ground testing. Two out of the three mounting interfaces to the heat-shield support panel (in the lab setup replaced by the Aluminium plate) can be seen. Note that the images of the HRT (and thus our wavefront plots) are rotated by $90^\circ$ with respect to the laboratory frame.  }
  
     \label{HREW}
\end{figure}

\begin{figure*}[!h]
\centering
\includegraphics[width=\textwidth]{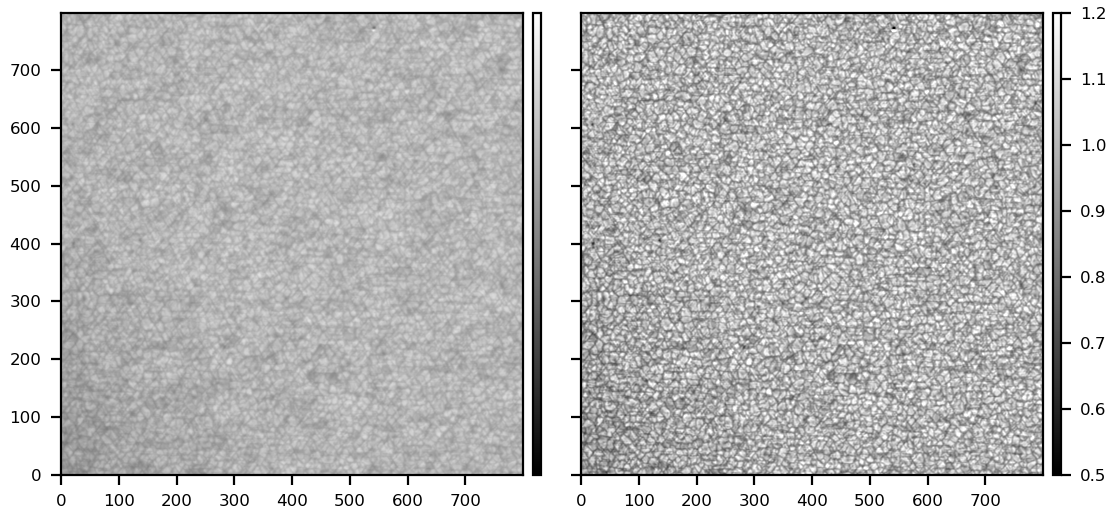}
  \caption{Reconstruction of the RSW2 data. The central region ($800\times800$ pixels) of the focused image of the PD pair of 22 March 2022 at $0.334$\,au (left panel). The reconstructed region with the PSF calculated by PD (right panel).
  }
  
     \label{raw_restored_foc}
\end{figure*}

\begin{figure}[!h]
\centering
\includegraphics[width=\columnwidth]{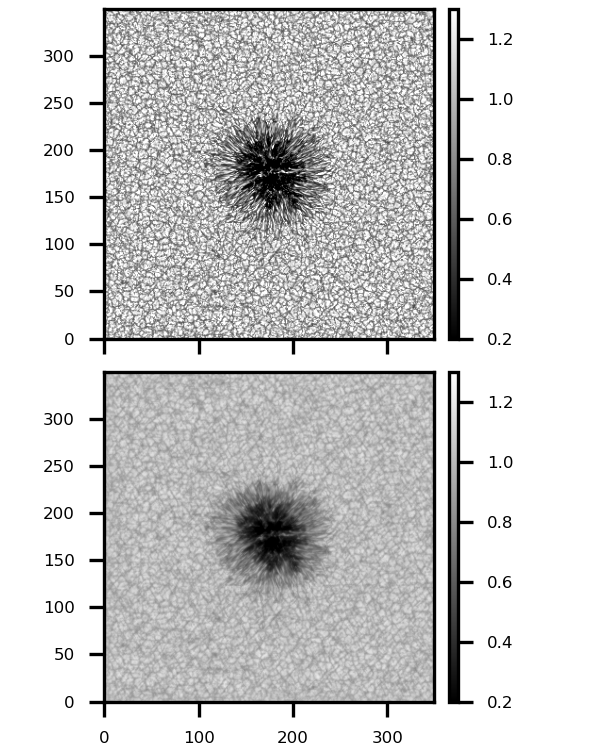}
  \caption{MURaM simulation of a sunspot. The original synthesised continuum map (upper panel), rebinned to the pixel size of HRT at $0.334$\,au, which is equal to $121$\,km. The rebinned and degraded map (lower panel) with the PSF calculated from the PD analysis of the PD image pair of 22 March 2022.
  }
  
     \label{muram_deg}
\end{figure}

\section{Phase Diversity analysis}
\label{pd_analysis}
We adopt the PD algorithm presented by \cite{loefdahl_scharmer_94}. The procedure for fitting the wavefront error (and all corresponding references) is described in \cite{kahil_2022_PD}. We use Noll's expansion scheme of orthonormal Zernike polynomials \citep{noll_zernike_1976} to characterise the wavefront error.

The PD pair is acquired during the second remote-sensing window (RSW2) on 22 March 2022 and at a distance of $0.334$\,au. The image pair is taken in the continuum of the SO/PHI spectral line. The artificial defocus introduced to the focused images was chosen to be half of the SO/PHI wavelength ($\lambda/2$). The PD image pair is shown in Figure~\ref{pd_pair_march}.
Before fitting the WFE, we align, using cross-correlation, the defocused image to sub-pixel accuracy to the focused image. We therefore disregard the first three aberrations (piston, tip, tilt) and start the WFE fitting from the fourth Zernike polynomial (defocus, $Z4$).

We run the PD algorithm on four sub-regions of the FOV, each of an equal size of $750\times750$ pixels. The optimal number of the employed Zernike polynomials which returns a valid WFE is $Z=23$ (from $Z4$ to $Z26$). Any number larger than $Z=23$ results in an over-fitted WFE and an over-reconstructed scene due to the noise amplification. The dependence of the WFE fitting and restoration results on the employed number of Zernike polynomials is discussed in \cite{hirzberger_performance_2011}. We show the results of the WFE fitting in Figure~\ref{wfs_mtfs_march}. As expected, the spatial variation of the WFE across the FOV is small and the images can be assumed to be isoplanatic, in agreement with earlier results of \cite{kahil_2022_PD}.

To retrieve the set of Zernike coefficients to be used for characterising the WFE of HRT during the RSW2 and for comparison with earlier results, we use the averaged Zernike terms over the four sub-regions of Figure~\ref{wfs_mtfs_march}. The Zernike coefficients are shown in the bar plot of Figure~\ref{all_zernikes}. Overplotted is the distribution of the Zernike coefficients retrieved during earlier PD measurements in NECP on 20 April 2021 ($0.82$\,au) and during CP on 20 February 2021 ($0.52$\,au). For these orbits, we employed only 10 Zernike polynomials. This number was chosen because the total root mean square (RMS) of the WFE saturates for $Z > 10$. We summarize the results of the PD analysis and compare them to earlier results in Table~\ref{tab1}.

 As expected, the Zernike terms that mostly increase in amplitude during the RSW2 are the defocus (Z4), the first order trefoil (Z9, Z10) and spherical aberration (Z11). This is a result of the deviation of the actual temperature distribution across the HREW from a paraboloid shape. These deviations give rise to spherical aberration, and to trefoil due to azimuthal inhomogeneities.

The contribution of the trefoil terms is apparent in the trigonal shape of the wavefront error in Figure~\ref{wfs_mtfs_march}. This shape is related to the mount of the HREW in the heat-shield of SO and attributed to a trigonal temperature inhomogeneity on the HREW. The inhomogeneity is  caused by heat conduction through the mount points of the HREW within the feed-through of the spacecraft heat-shield (which were designed to minimise thermal coupling), and of the hot feed-through to the heat-shield support panel, which sees the strongest temperature difference (see Figure~\ref{HREW}). 

Figure~\ref{raw_restored_foc} shows a sub-region of the focused image of the PD pair acquired on 22 March 2022 at a solar distance of $0.334$\,au. 
The trigonal pattern on top of the solar scene can be observed (left panel). We construct the PSF of HRT from the calculated best-fit Zernike polynomials and apply image restoration through deconvolution with the Wiener filter. The restoration (right panel) successfully removes the trigonal pattern caused by the HREW.

To test if the PSF deduced from the PD analysis of the pair taken during RSW2 represents the true aberrations emanating from the HREW of HRT, we degrade a synthesised continuum image obtained using MURaM \citep{vogler_2005} with the calculated HRT PSF. The degradation is applied to the theoretical image, which is then rebinned to the pixel size of HRT after adding Gaussian noise (upper panel of Figure~\ref{muram_deg}). The degraded image (lower panel of Figure~\ref{muram_deg}) displays the same trigonal pattern as seen in the degraded HRT data of the RSW2 (left panel of Figure~\ref{raw_restored_foc}). This indicates that the wavefront fitting algorithm returns a reasonable set of Zernike coefficients that describe the true aberrations produced by the HREW of HRT.

\begin{figure}[!h]
\includegraphics[width=\columnwidth]{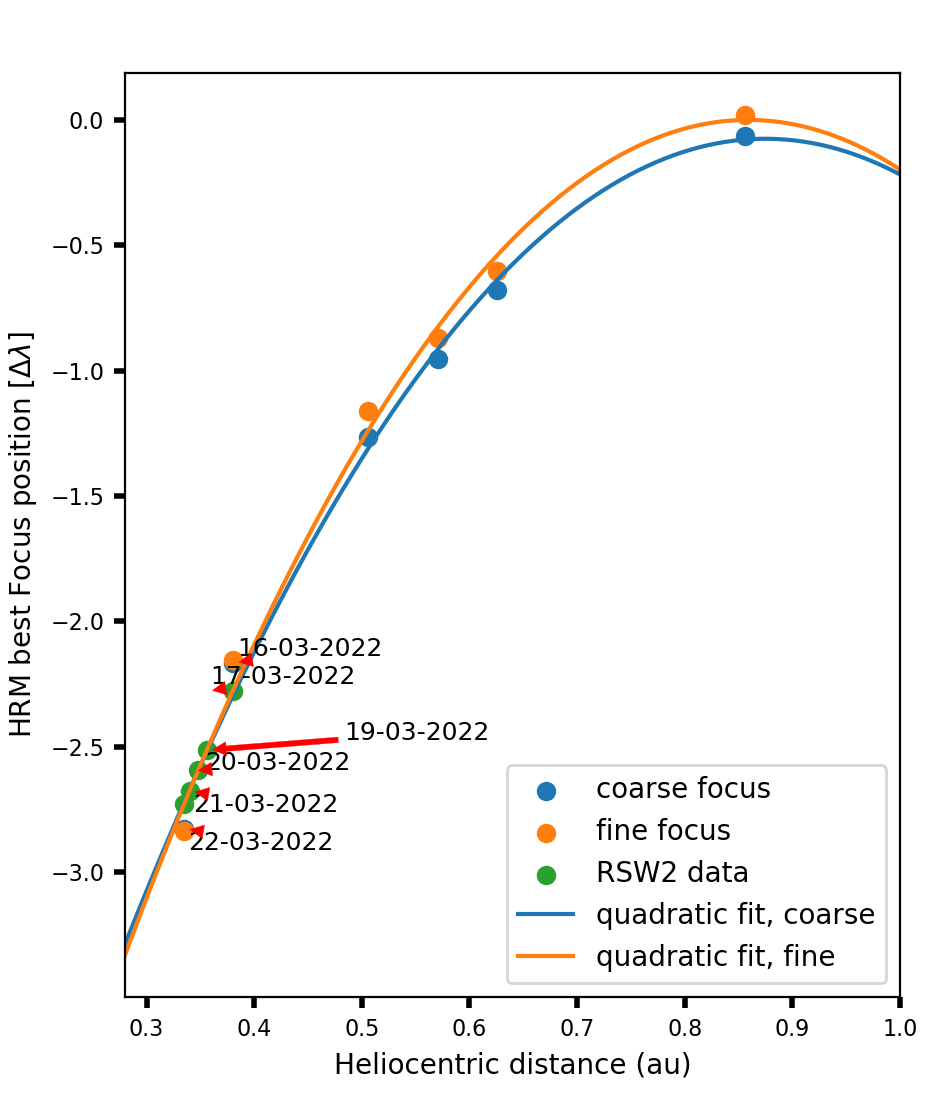}
  \caption{The variation of the HRM best focus position in coarse and fine focus (in units of $\lambda$) along the orbit of SO from 05 November 2021 to 22 March 2022. The curves correspond to the best-fit quadratic function to both types of focus. The y-scaling is chosen such that the HRM best focus position at the absolute maximum of the quadratic fit is equal to zero.} 
  
     \label{hrm_vs_d}
\end{figure}
\begin{figure}[!h]
\centering
\includegraphics[width=\columnwidth]{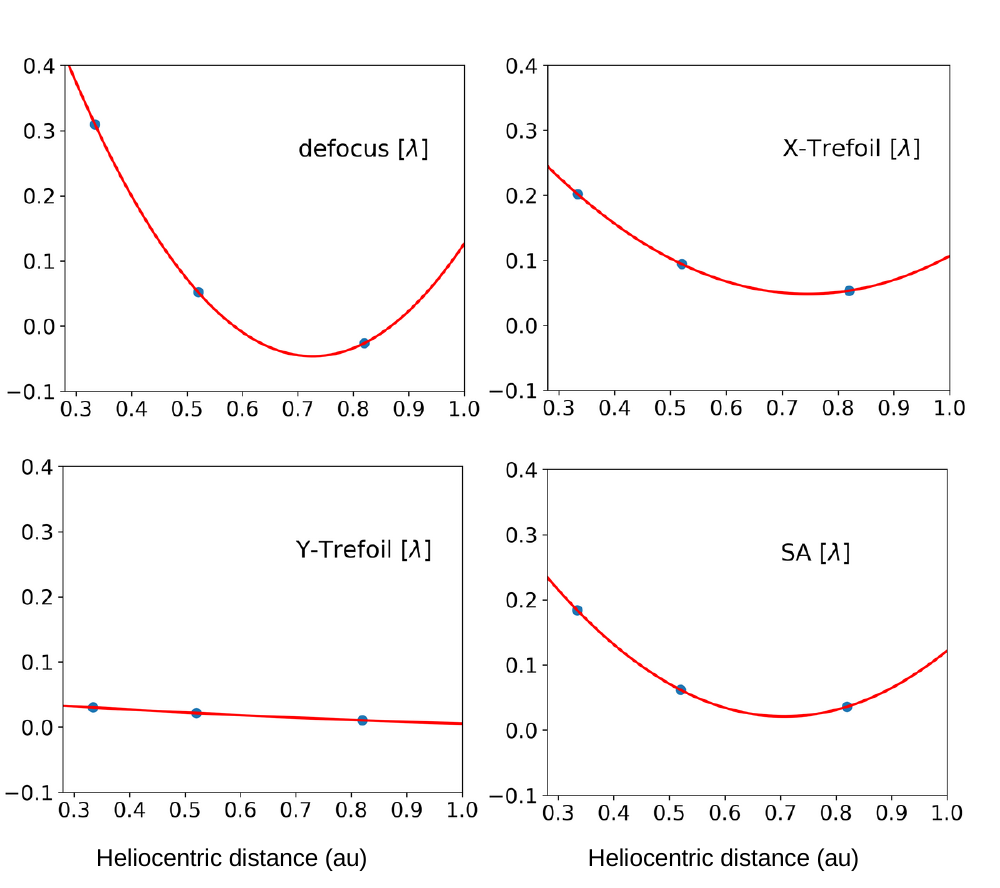}
  \caption{The variation of the four Zernike coefficients (defocus, spherical aberration, and the two components of 1st order trefoil) in units of $\lambda$ with distance of SO to the Sun (in au). The red curves are the quadratic fits to the data points. The absolute value of the X-Trefoil aberration is plotted in order to use the same scaling on the y-axis .}
  
     \label{fit_zernikes}
\end{figure}

\begin{figure}[!h]
\centering
\includegraphics[width=\columnwidth]{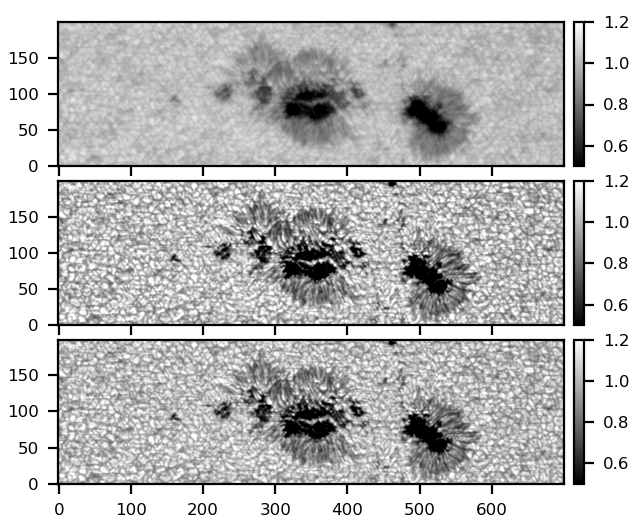}
  \caption{Restoration with the interpolated PSF. The continuum image of a sub-region of one dataset from 17 March 2022 at $0.379$\,au (upper panel). The restored region with the PSF calculated at $0.334$\,au (middle panel). The restored region with the PSF interpolated to a distance of $0.379$\,au (lower panel).}
  
     \label{march17_psf_march22}
\end{figure}

\begin{figure*}[!h]
\centering
\includegraphics[width=\textwidth]{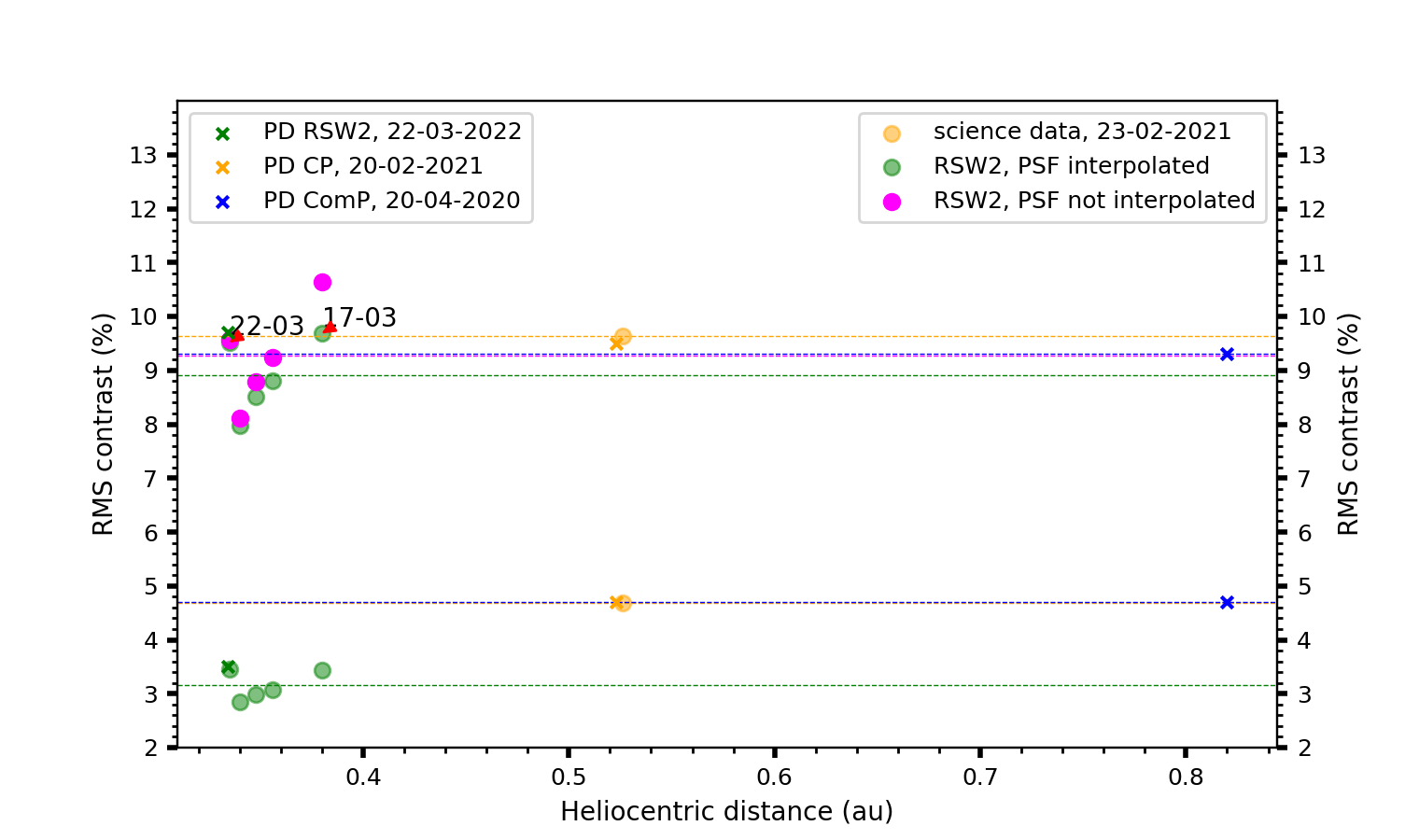}
  \caption{The variation of the initial (lower data points) and reconstructed (upper data points) contrast values in the SO/PHI continuum along the orbit of SO. The dots correspond to science data while the crosses correspond to the PD datasets. Each data point of the RSW2 data (green and magenta dots) is the average of all daily observations on 17, 19, 20, 21 and 22 March 2022. The dashed horizontal lines represent the averaged contrasts of all the points/crosses of the corresponding colour. }
  
     \label{all_contrasts_orbits}
\end{figure*}

\section{PSF interpolation}
\label{psf_interpolation}
The calculated PSF was estimated at a solar distance of $0.334$~au. Since the aberrations introduced by the HREW are expected to increase with decreasing solar distance of SO (see Figure~\ref{all_zernikes}), the HRT data obtained at different orbital positions cannot be restored using the same calculated PSF. This will result in an over-reconstruction for data recorded when SO was further from the Sun than $0.334$\,au. Therefore, we develop an interpolation scheme to approximate the PSF for such distances where no PD image pairs were acquired. 

We assume that the instrument does not change with orbital position and only the temperature inhomogeneities on the HREW vary quadratically with solar distance. This behaviour is apparent in the plot of the total WFE compensated by the HRM along the orbit shown in Figure~\ref{hrm_vs_d}. 

For constructing the interpolated PSF at a given distance below $0.5$\,au (where artifacts due to the HREW are significant), we use the 23 Zernike coefficients deduced from the analysis of the RSW2 PD data at $0.334$\,au (orange bars in Figure~\ref{all_zernikes}) and interpolate with SO-Sun distance only the terms which are mostly affected by the temperature of the HREW. These are: defocus ($Z4$), first order trefoil components ($Z9$, $Z10$) and spherical aberration ($Z11$). We use a quadratic function to model the variation of these aberrations with distance. The choice of a quadratic function is motivated by the following arguments: (1) The temperature across the HREW goes quadratically with distance, and the WFE is temperature dependent; (2) the HRM compensates for the defocus in HRT following a quadratic dependence on the heliocentric distance (see Figure~\ref{hrm_vs_d}); (3) a quadratic function is the highest order that could be fit unambiguously to three data points. The results of the fits are shown in Figure~\ref{fit_zernikes}. The interpolated values are plotted for the range of 0.28\,au (the closest perihelion of SO) to 1\,au. We note that we are only interested in reconstructing data from distances below 0.5\,au, where the HREW causes a significant reduction of the image quality. The HRT data taken at distances larger than 0.5\,au are corrected with the PSF calculated during the cruise phase at 0.5\,au (see Section~\ref{rsw1_correction}) .
 
We show in Figure~\ref{march17_psf_march22} a sub-region of a dataset taken on 17 March 2022 in the SO/PHI continuum. The dataset was acquired at a solar distance of $0.379$\,au. The restoration is done with both, the PSF calculated at $0.334$\,au and the PSF interpolated to a distance of $0.379$\,au. The aberrations due to the HREW appear to be greatly reduced in both images, but without PSF interpolation, images appear to be over-reconstructed. We quantify this effect by calculating the normalized RMS contrast of the continuum intensity of quiet-Sun regions in the RSW2 observations before restoration, after restoration with the PSF determined at 0.334\,au and after restoration with the interpolated PSF. The contrast values are plotted in Figure~\ref{all_contrasts_orbits}. For comparison, we also show the contrast values of reconstructed scenes from earlier orbits. These correspond to the focused images of the PD datasets, and they are marked as crosses in Figure~\ref{all_contrasts_orbits}.

The PD datasets are acquired by HRT shortly after refocusing with the HRM, so that the obtained contrast of 9.3\% to 9.7\% in the corresponding reconstructed solar scene are considered to be optimal. During the perihelion approach the overall WFE is deteriorating rapidly, as depicted in Figure~\ref{hrm_vs_d}. Since re-focusing with the HRM was not done between 16 March 2022 and 22 March 2022 (see Figure~\ref{hrm_vs_d}), we witness a significant decrease in contrast in the corresponding initial images (lower green dots). Deconvolution with the interpolated PSF (upper green dots) does increase the contrast significantly, but the contrast values in these datapoints do not fully reach the optimum value, which is only re-established once the system has been brought to optimum focus again by the HRM (on 22 March 2022).

Using the nearest PSF estimated on 22 March 2022 for all data points (magenta dots) shows clearly, that -- with the natural exception of the March 22 point itself -- all data points suffer from over-reconstruction, since the applied PSF corresponds to the worst WFE, while the data have been taken with a more relaxed instrument.



\begin{figure}
\centering
\includegraphics[width=\linewidth]{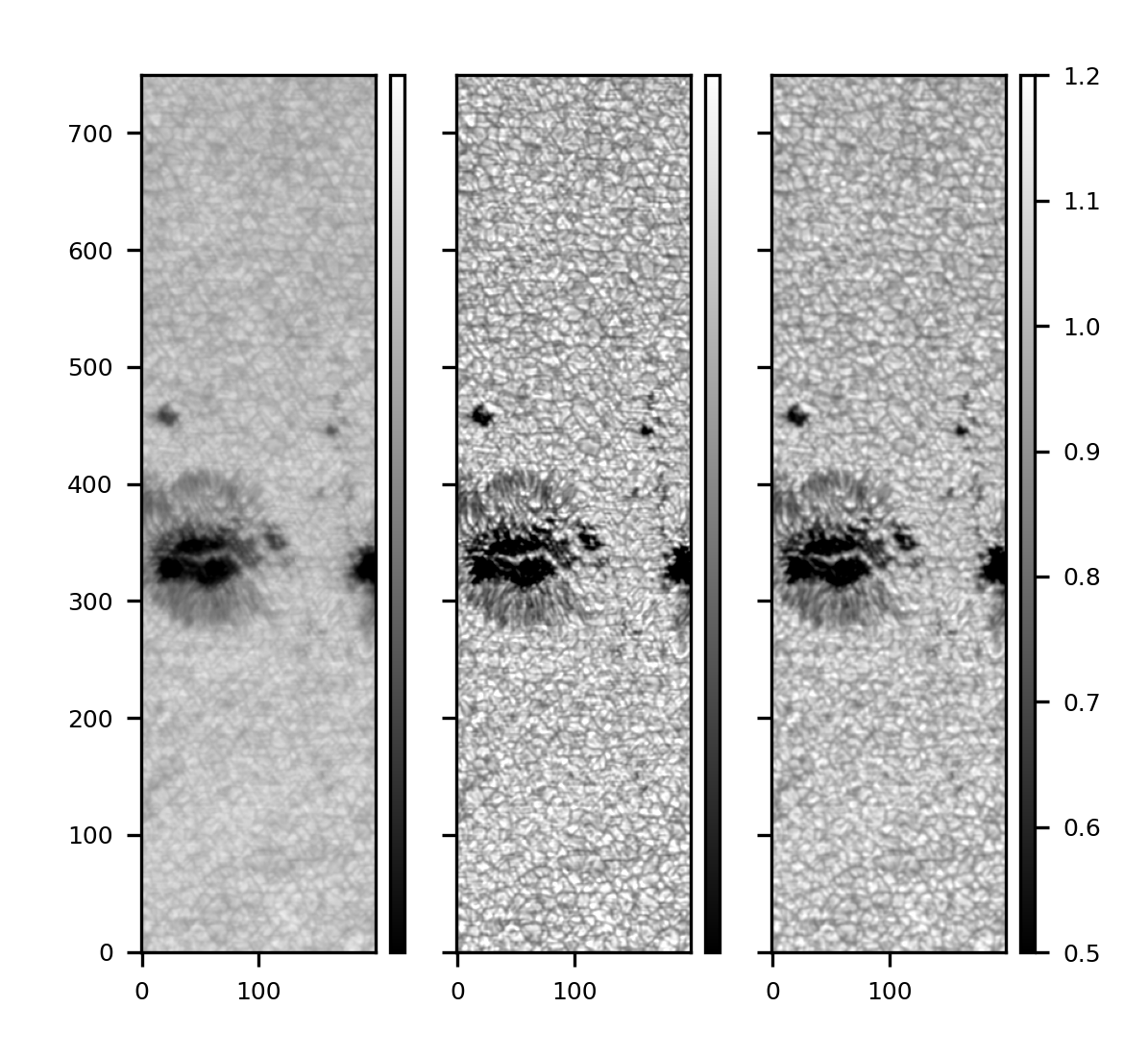}
  \caption{Results of the aberration correction. A sub-region of an example dataset taken on 17 March 2022 at 03:18:09 UT (first panel). The restored area with the interpolated PSF (middle panel) and the aberration corrected area (last panel).
  The maps are taken in the continuum of the SO/PHI spectral line at $617.3$\,nm. }
  
     \label{results_cont}
\end{figure}
\begin{figure}
\centering
\includegraphics[width=\linewidth]{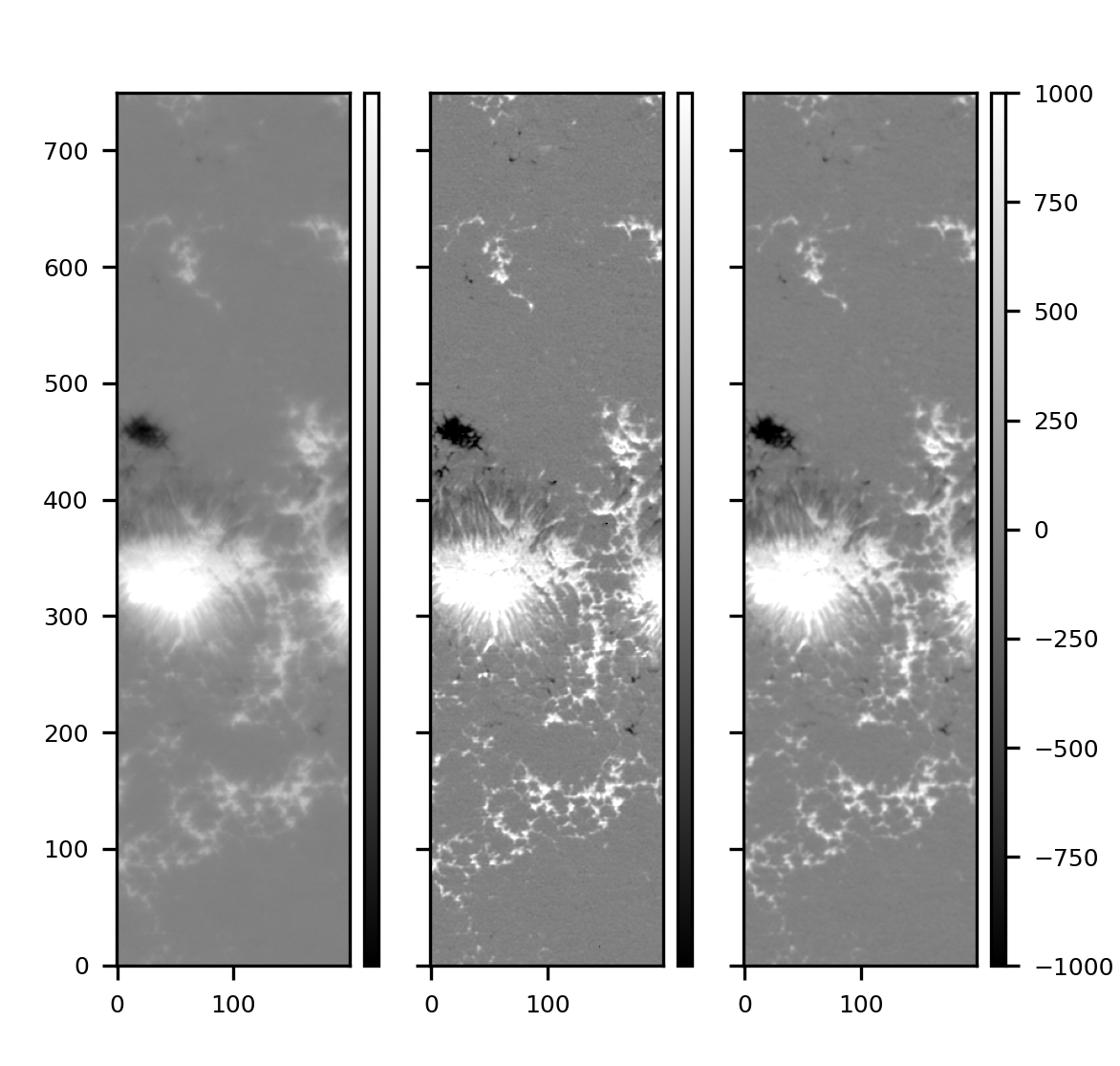}
  \caption{Same as Figure~\ref{results_cont}, but for the LOS magnetograms obtained from the inversions of the reconstructed Stokes maps.}
  
     \label{results_blos}
\end{figure}

\section{Aberration correction}
\label{aberration_correction}

\subsection{RSW2 data}

The restoration of the full Stokes images with the interpolated PSF results in amplified noise levels compared to non-restored data. This is an expected drawback of deconvolution, which cannot be avoided \citep[see discussion in][]{kahil_2022_PD}. Usual PD reconstructions are known to increase the noise in the restored data by a factor of three with respect to the original degraded data \citep[see for example][]{imax_2011}. Therefore, the polarisation signals of small-scale magnetic structures in the restored data may lie within the noise level, which is not convenient for studying such structures.  

To reduce the noise level, while removing the aberrations caused by the HREW, we convolve the restored images with the PSF of an ideal telescope of 140\,mm aperture, i.e. by an Airy function with its first zero at the diffraction limit. This results in lower noise amplification at higher spatial frequencies due to deconvolution. In this work, the term "restored data" refers to the reconstructed data with the corresponding interpolated PSF, and with a noise level equal to roughly 3$\sigma$ ($\sigma$ being the noise level in the non-restored data,which is estimated from the distribution of Stokes~$V$ in the continuum and of the LOS magnetic field ($\rm B_{\rm los}$) signals in quiet-Sun areas, as described below), while "aberration-corrected data" refers to data corrected only for the aberrations caused by the HREW while keeping the degradation caused by diffraction.

The restoration and aberration corrections are applied to the individual Stokes images before the Milne-Eddington inversion of the radiative transfer equation is performed. This step is being implemented in the on-ground HRT data reduction pipeline presented in \cite{sinjan_pipeline}. Example studies which make use of the aberration-corrected HRT data have been carried out by \cite{calchetti_2023} and \cite{sinjan_2023}. We show in Figures~\ref{results_cont} and \ref{results_blos} sub-regions of the continuum and $\rm B_{\rm los}$ maps of an example dataset. The figures show the two levels of correction with the HRT PSF. The first panels of each figure correspond to the non-restored data, while the panels in the middle correspond to the restored version with noise levels of about $3\sigma$. The last panels are the aberration-corrected data with lower noise levels. We calculate the amount of noise reduction in the following subsections.

\begin{figure}[!h]
\includegraphics[width=\columnwidth]{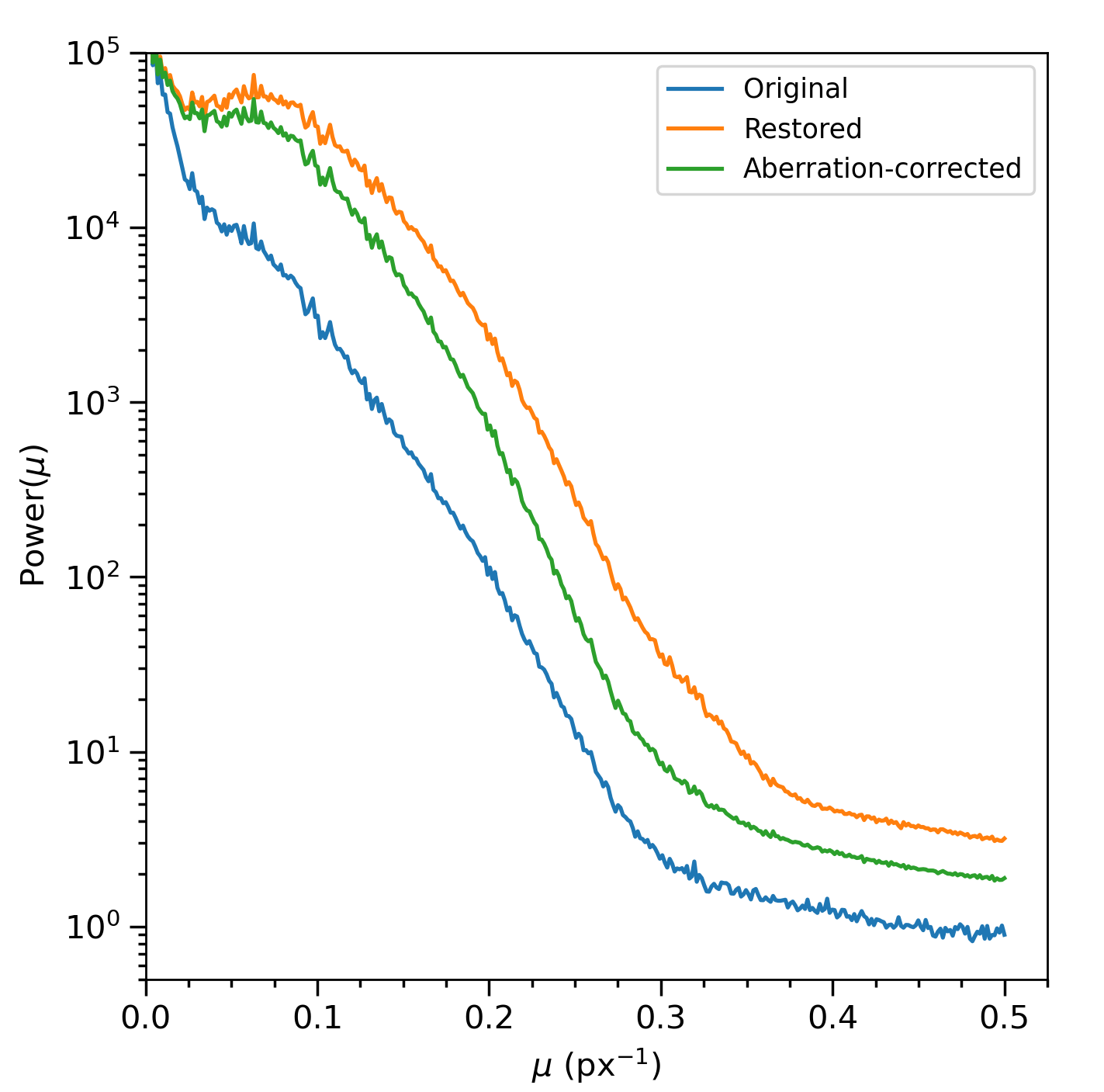}
  \caption{Power spectrum of the continuum intensity in a quiet-Sun region of a dataset taken on 17 March 2022 for the two levels of data reconstruction (data restoration and aberration correction). }
  
     \label{1d_ps}
\end{figure}

We illustrate in Figure~\ref{1d_ps} the one dimensional ($1$D) power spectrum of a quiet-Sun region taken on 17 March 2022 at 00:20:09 UT. The three curves correspond to the same region in the original, restored and aberration-corrected versions. The plot shows that the aberration correction lowers the signal towards higher frequencies, which reduces the noise in the aberration corrected data. It also lowers the signal at intermediate frequencies, which results in lower image contrast: the RMS contrast of quiet-sun regions decreases, on average, from 9\% to 7\% due to the aberration correction (the initial contrast is equal to 3.4\%).

\begin{figure*}[!h]
\centering
\includegraphics[width=16.2cm]{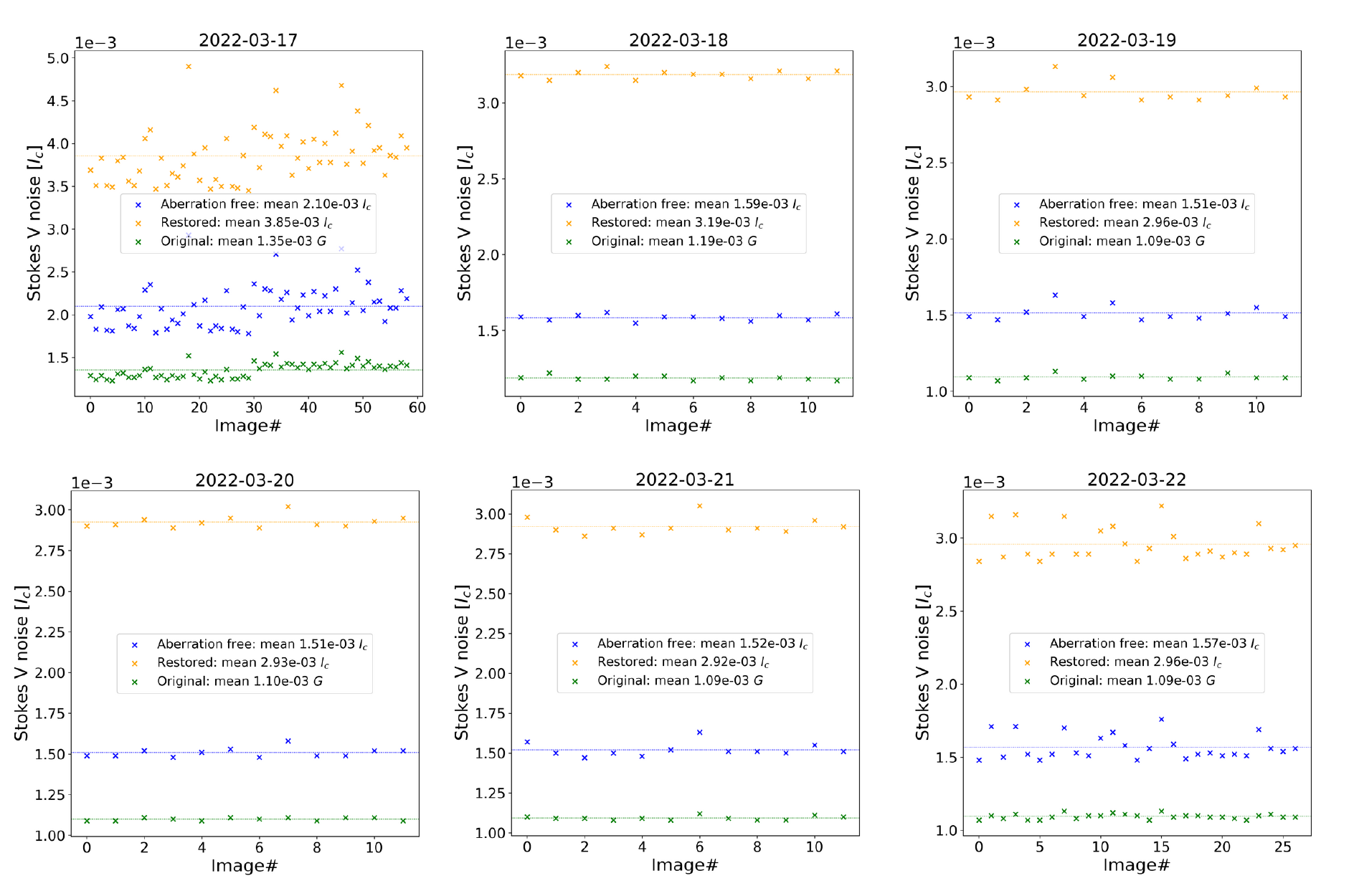}
  \caption{Noise as measured in the Stokes~$V$ continuum of the HRT spectral line (in units of mean quiet-Sun intensity). The noise is measured in every dataset of the RSW2 observations and for the three data levels: L2 degraded data (green), restored with the interpolated PSF L2 data (orange), and aberration-corrected L2 data (blue). The box in each plot displays the computed mean of the measured noise over all daily observations.  }
  
     \label{noise_stokesV_rsw2}
\end{figure*}

\begin{figure*}
\centering
\includegraphics[width=16.2cm]{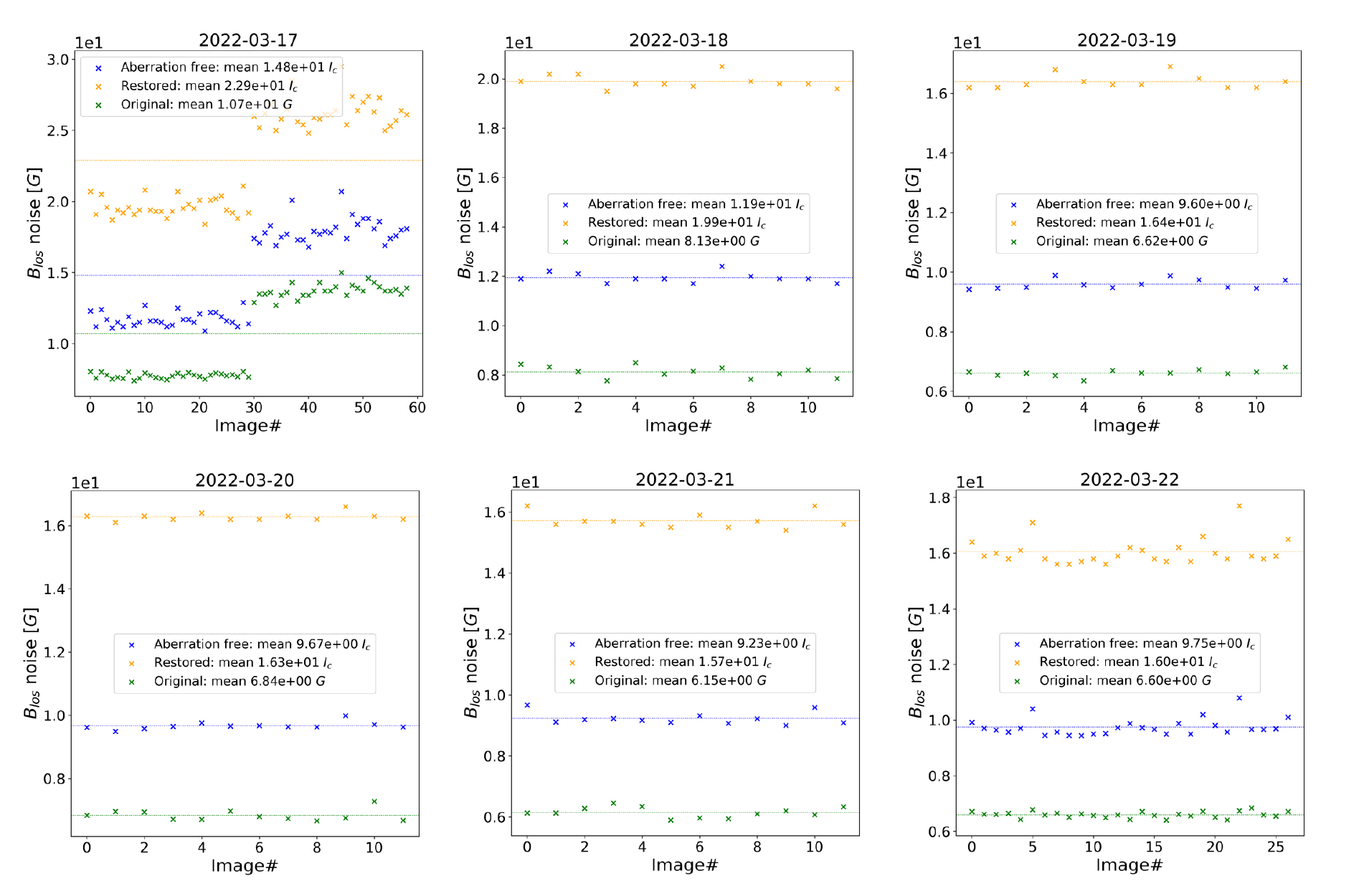}
  \caption{Same as Figure~\ref{noise_stokesV_rsw2} but for the noise measured in the LOS magnetic field maps of the RSW2 datasets. The noise is measured from Gaussian fits to the histograms of the $\rm B_{los}$ distribution as described in \cite{Liu_2012}}.
  
     \label{noise_blos_rsw2}
\end{figure*}

To quantify the amount of noise reduction, we measure the distribution of Stokes~$V$ signals in the continuum in quiet-Sun areas. We also follow the method in \cite{Liu_2012} to compute the noise in  the LOS magnetic field maps. The results are shown in Figures~\ref{noise_stokesV_rsw2} and \ref{noise_blos_rsw2} where the noise is measured for all datasets of the RSW2 observations.
Averaged over all days of the RSW2 observations, the noise in Stokes~$V$ increases by a factor of only $1.45$ in the aberration corrected data compared to the original data. This factor is equal to $3$ upon full restoration of the data.

\subsection{RSW1 data}
\label{rsw1_correction}
\begin{figure*}
\centering
\includegraphics[width=\textwidth]{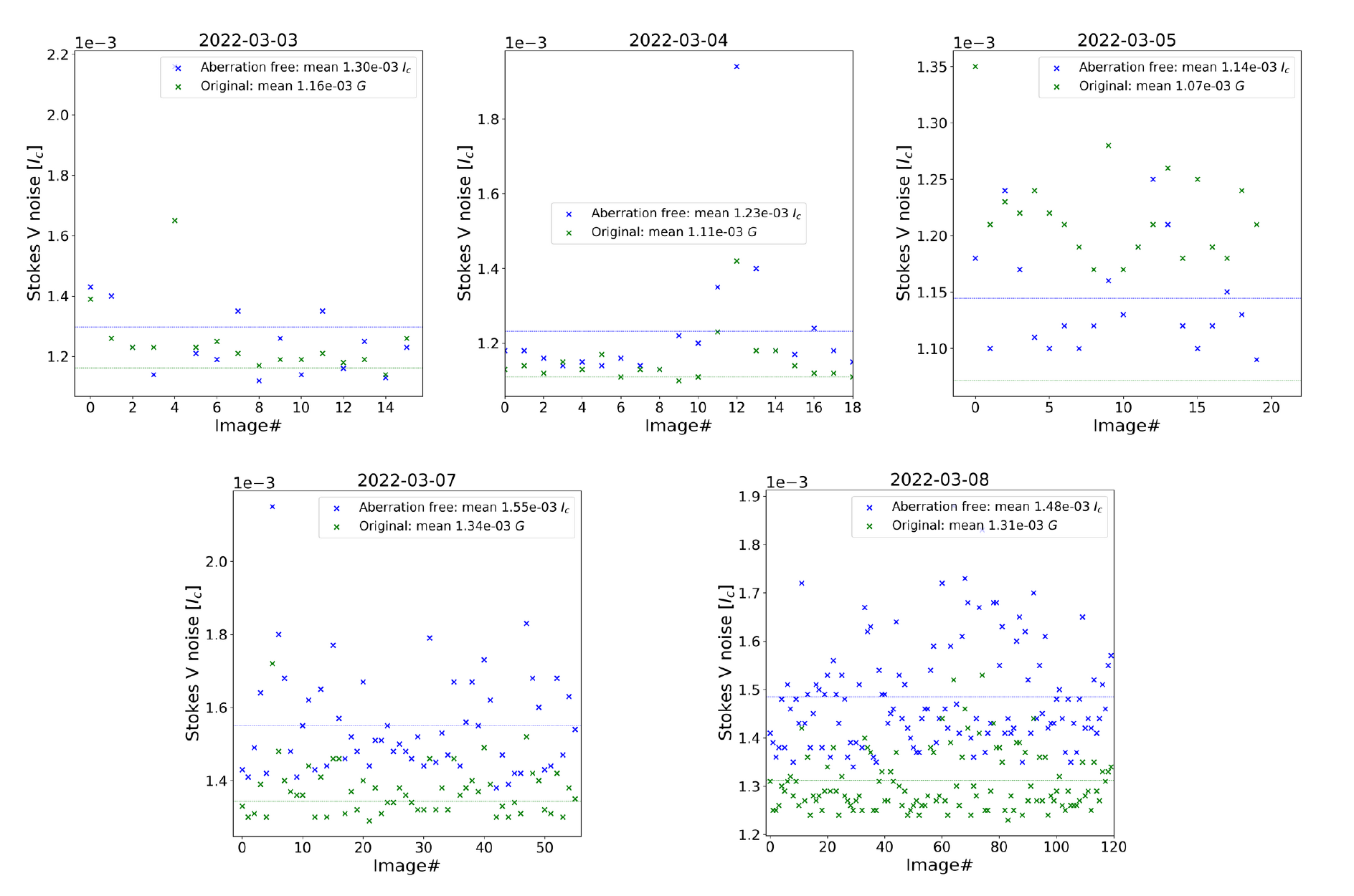}
  \caption{The noise levels in the original (green) and aberration-corrected data (blue) of the RSW1 observations. The noise is measured in the continuum of Stokes~$V$ for each of the datasets. The restored data (with 3$\sigma$ noise level) are not fully available, so we do not show the corresponding noise values.}
  
     \label{noise_stokesV_rsw1}
\end{figure*}

\begin{figure*}
\centering
\includegraphics[width=\textwidth]{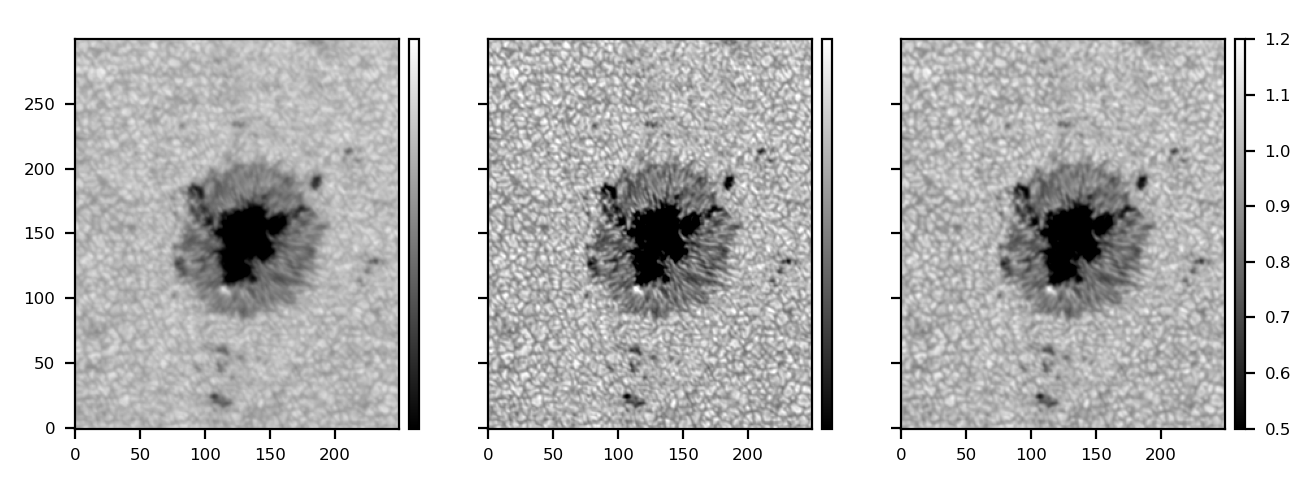} 
  \caption{The restored and aberration-corrected RSW1 data. A sub-region of an example dataset taken on 03 March 2022 at 09:40:09 UT in the continuum (left panel). The restored region with the PSF estimated during the cruise phase (middle panel) and the aberration-corrected region (right panel).The SO-Sun distance was $0.547$\,au  }
  
     \label{restoration_march_03_cont}
\end{figure*}

\begin{figure*}[!h]
\centering
\includegraphics[width=\textwidth]{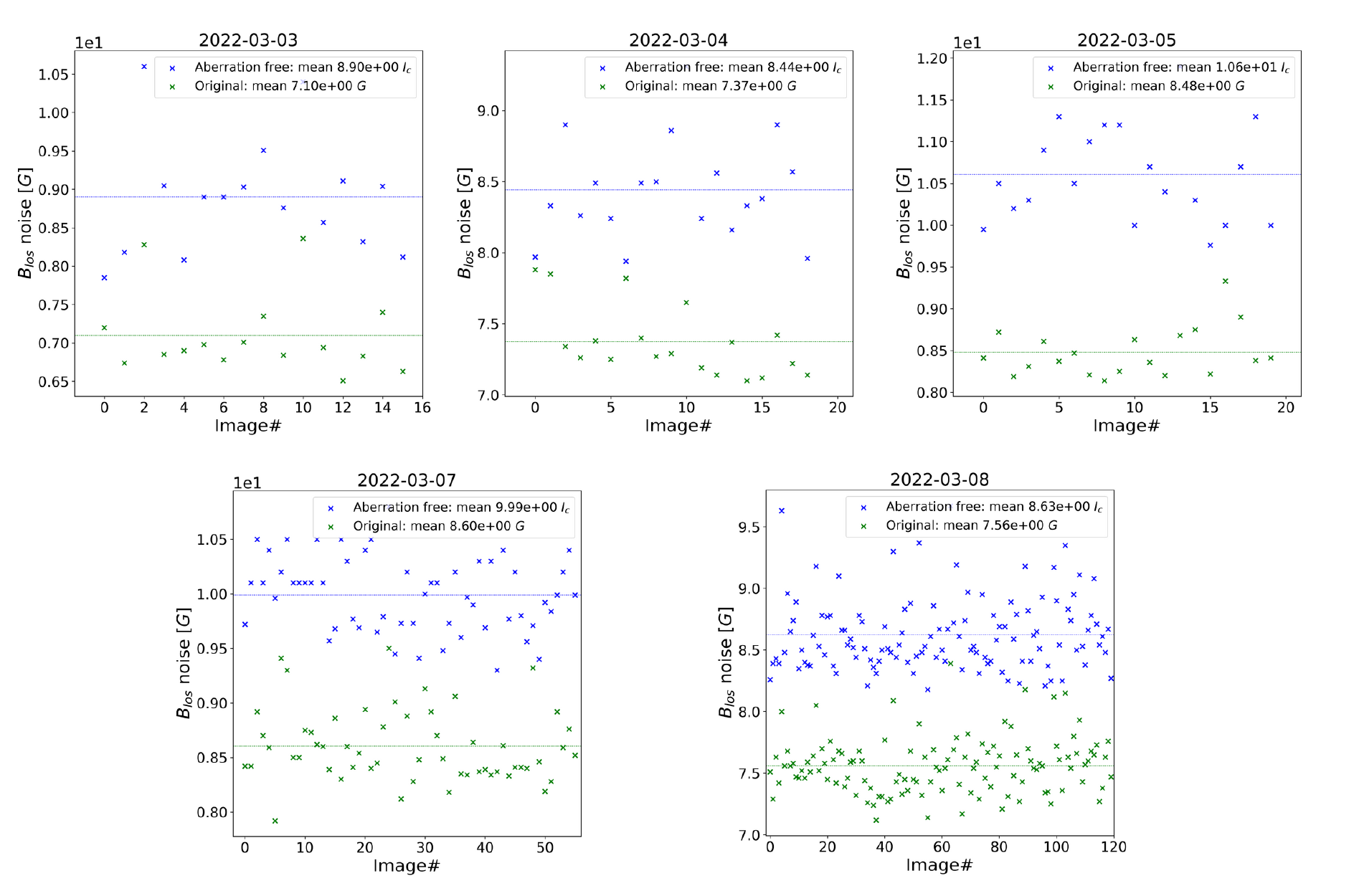}
  \caption{ The noise in the LOS magnetograms of the aberration-corrected data of the RSW1 observations.}
  
     \label{noise_blos_rsw1}
\end{figure*}

\begin{figure*}
\centering
\includegraphics[width=\textwidth]{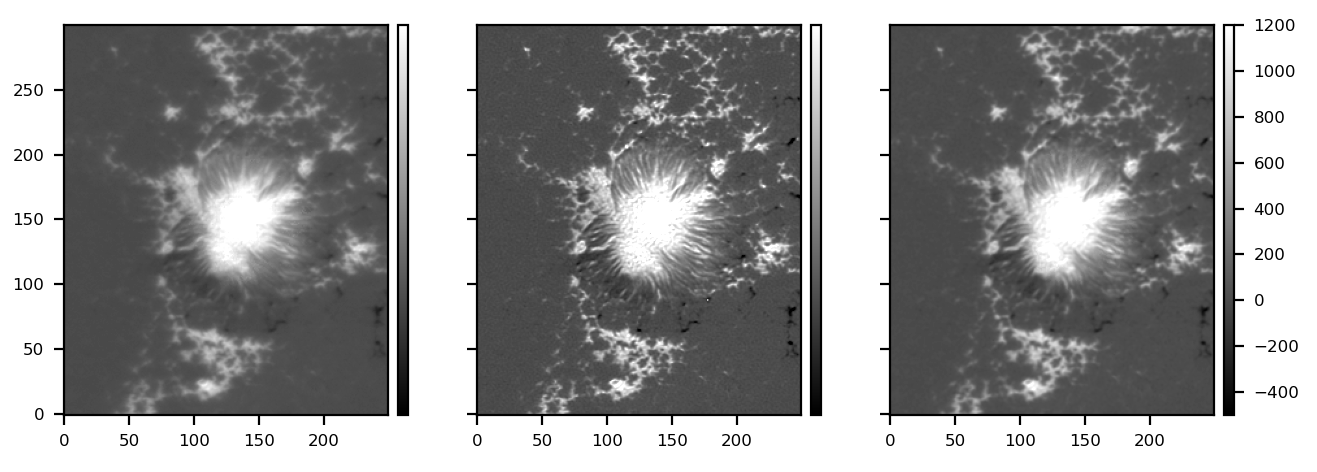}
  \caption{Same as Figure~\ref{restoration_march_03_cont} but for the LOS magnetogram.}
  
     \label{restoration_march03_blos}
\end{figure*}

During the first remote sensing window of SO (RSW1), the HRT datasets were acquired at heliocentric distances ranging from $0.547$\,au to $0.489$\,au. Therefore, the degradation in these data by the HREW is smaller than for the RSW2 data described earlier. However, for consistency with the released RSW2 data, we also apply the aberration correction. No PD measurements were acquired during RSW1 so we use the PSF estimated during the cruise phase at a similar distance from the Sun of $0.5$\,au. This PSF had a smaller RMS WFE and lower number of Zernike polynomials (see Table~\ref{tab1}), so that the aberration correction results in increased noise levels of only a factor of $1.2$ wrt non-restored RSW1 data (see Figures~\ref{noise_stokesV_rsw1} and \ref{noise_blos_rsw1}). We show in Figures~\ref{restoration_march_03_cont} and \ref{restoration_march03_blos} an example of one HRT dataset taken on 03 March 2022 at a distance of $0.547$\,au. We show the restored image and the aberration-corrected one. 
\section{Discussion and conclusions}
\label{discussion}
In Sections~\ref{pd_analysis} and \ref{psf_interpolation} we have calculated, by means of phase diversity analysis, the wavefront error of HRT when close to the Sun during the second remote sensing window of the nominal phase mission. At such distances, the breaking of the rotational symmetry of the temperature gradients across the HREW introduces aberrations which  cannot  be corrected for by the HRT refocus mechanism, but are well captured by PD calibration data. 

By inspecting the variation of the Zernike terms describing the WFE with solar distance, we could isolate the coefficients which are introduced by this effect (defocus, trefoil and spherical aberration). The fact that we still see a defocus term, even after refocusing by the HRM, can be explained by the action of the autofocus system, which aims to optimize the RMS contrast in the image by minimising the overall wavefront error. But, since the shape of the wavefront is not parabolic, there is a trade-off between the rotationally symmetric terms, mainly defocus and spherical aberration, such that after the "refocusing" we see remaining contributions of both of these terms.

With the set of PD data available so far, we build an interpolation model for the calculated Zernike coefficients in order to construct a PSF that depends on the heliocentric distance. This PSF is employed to reconstruct the corresponding HRT datasets close to perihelion ($<0.5$\,au), when no PD measurements are available. We showed that the deconvolution with an interpolated PSF yields very good reconstruction of the original scene, as long as the instrument has been brought to best focus by the HRM and the residual WFE has been minimized. Refocusing by the HRM on a regular basis especially during perihelion approach, where the temperature changes rapidly is
highly recommended in order to minimize the overall WFE prior to PSF deconvolution. Under these preconditions, the smooth variation of the residual WFE terms allows good reconstruction of the solar scene, without strong amplification of noise, which would happen when the initial contrast gets too low due to insufficient focusing. 

In addition, we have proposed, in Section~\ref{aberration_correction}, a solution to avoid the noise amplification which results from the reconstruction with the interpolated PSF. The resulting data products are termed "aberration corrected" with a noise level that is on average $1.45$ larger than the noise in the non-reconstructed RSW2 data.  For RSW1 data where the degradation effect by the window is smaller, the noise of the aberration corrected data increases by a factor of $1.2$. This moderate increase of the noise level makes the aberration-corrected data ideal for applying inversion methods, whereas studies based on the intensity images might benefit from the full interpolated PSF restoration.

For future near-perihelion observations, HRT will acquire daily PD measurements in order to study in more detail the variation of the aberrations introduced by the HREW with solar distance, and with different pointings of the spacecraft. In addition to these measurements, potential long-term effects, which are not a direct function of distance alone cannot be ruled out at the current time and they need to be further investigated in a future study. Furthermore, acquiring multiple defocused datasets may increase the accuracy of the wavefront error retrieval by the PD algorithm, as shown by \cite{bailen_2023}.

\begin{acknowledgements}
Solar Orbiter is a space mission of international collaboration
between ESA and NASA, operated by ESA.  We are grateful to the ESA SOC
and MOC teams for their support.  The German contribution to SO/PHI is
funded by the BMWi through DLR and by MPG central funds. The Spanish
contribution is funded by FEDER/AEI/MCIU (RTI2018-096886-C5), a
“Center of Excellence Severo Ochoa” award to IAA-CSIC
(SEV-2017-0709), and a Ramón y Cajal fellowship awarded to DOS. The
French contribution is funded by CNES.
\end{acknowledgements}

\newpage
\bibliographystyle{aa}
\bibliography{solo_paper_psf}


\end{document}